\documentclass[12pt,preprint]{aastex}

\slugcomment{17 Apr 2002 Submitted}
\shorttitle{Merging of Brightest Cluster Galaxies}
\shortauthors{Yamada et al.}
\begin{document}
%
%
\title{Witnessing the Hierarchical Assembly of the Brightest Cluster Galaxy
in a Cluster at z=1.26}
\author{Toru \textsc{Yamada}\altaffilmark{1}}%
\email{yamada@optik.mtk.nao.ac.jp}
\author{Yohei \textsc{Koyama},\altaffilmark{2}
        Fumiaki {\sc Nakata},\altaffilmark{3} 
        Masaru {\sc Kajisawa},\altaffilmark{1}
        Ichi {\sc Tanaka},\altaffilmark{1} \\
        Tadayuki {\sc Kodama},\altaffilmark{3}
        Sadanori {\sc Okamura},\altaffilmark{3} 
        and 
        Roberto {\sc De Propris},\altaffilmark{4} \\}
\altaffiltext{1}{National Astronomical Observatry of Japan, 
       2-21-1, Osawa, Mitaka, Tokyo, 181-8588 }
\altaffiltext{2}{Astronomical Institute, Tohoku University, 
       Aoba-ku, Sendai, 980-8578 }
\altaffiltext{3}{Department of Astronomy, School of Science, The University of Tokyo, Bunkyo-ku, Tokyo 113-0033 }
\altaffiltext{4}{Research School of Astronomy and Astrophysics, Australian National University, \\
Weston, ACT, 2611, Australia
}

\begin{abstract}
  We have obtained a new high-resolution $K^\prime$-band image of the central region of the rich X-ray cluster RX J0848.9+4452 at z=1.26. We found that the brightest cluster galaxy (BCG) in the cluster is clearly separated into two distinct objects. Whereas the optical to near-infrared colors of the objects are consistent with the predictions of passive evolution models for galaxies formed at high redshift, the luminosities of the two galaxies are both considerably fainter than predicted by passive evolution of BCG's in low and intermediate redshift clusters. We argue that this is evidence of an on-going merger of normal cluster ellipticals to form the dominant galaxy in the core of RX J0848.9+4452. The two galaxies appear to point towards the nearby cluster ClG J0848+4453 \cite{sta97} and are aligned with the outer X-ray contour of their parent cluster, supporting a model of BCG formation by collimated infall along the surrounding large-scale structure.
\end{abstract}
 
\keywords{galaxies: elliptical and lenticular, cD---galaxies: evolution---galaxies: formation}

%
%
%
\clearpage
\section{Introduction}

 The cores of rich clusters are often dominated by massive elliptical galaxies: many of these brightest cluster galaxies (BCG's) are conspicuously different from other cluster members, in being at least one magnitude brighter than other giant ellipticals (Sandage 1976, Hoessel et al. 1980) and being located at or close to the bottom of the cluster potential well (Oegerle \& Hill 2001). It is likely that BCG's arise from merging of massive satellites, drawn towards the cluster center by dynamical friction, as also hinted by the high frequency of multiple nucleation in these objects (e.g., Furuzawa et al. 1999 and  references therein), or by collimated infall along filaments in the early epochs of cluster formation (West 1994; Dubinski 1998).

The $K$-band Hubble diagram provides a measure of BCG evolution but previous results have been contradictory: \cite{aes98} argued that BCG's dim with redshift according to, or fainter than, no-evolution models, suggesting that their mass has therefore increased in recent times by galaxy merging or cannibalism. On the other hand, \cite{com98} found evidence that BGC's evolve passively in an X-ray selected sample of clusters. \cite{bur00} suggested that the discrepancy is due to sample selection, as Aragon-Salamanca et al's (1998) sample consists of optically selected clusters some of which are not yet virialized enough to show strong X-ray emission.  Recently, \cite{nel01, nel02a, nel02b} have confirmed that (i) BCG's in different environments evolve differently \citep{bur00} and (ii) there has been significant growth in BCG mass in low X-ray luminosity clusters since at least $z \sim 1$.  However, \cite{bro02} recently found that there is no notable difference in the luminosity of brightest cluster galaxies between clusters with high and low X-ray luminosity at $z < 0.1$, suggesting that BCG evolution is nearly complete at the present epoch in cluster environments.

 Observations at high redshift allow us to study systems in an earlier stage of dynamical evolution and at a younger age. Among the small sample of $z > 1$ clusters discovered so far, RX J0848.9+4452 \citep{ros99} is one of the most secure examples of a rich, high concentration system resembling rich Abell clusters today. \cite{sta01} measured the temperature of the X-ray hot gas as T=5.8 keV and showed the gas has a regular, centrally concentrated spatial distribution. The central galaxy of the cluster is indeed the highest-redshift one in the sample studied in \cite{bur00}, and therefore provides the strongest leverage on the evolution of BCG's.

 We have exploited the good image quality achievable on the Subaru telescope to obtain very deep near-infrared images of this cluster at high spatial resolution. We find that the BCG is clearly resolved into two components in our $K^\prime$ image and interpret this result in the light of the $K$-band Hubble diagram of BCG's.

 In this paper, we focus on the properties of the BCG in the cluster core while our study of the luminosity function and color-magnitude diagram will be presented in a separate paper (Yamada et al. in preparation). We adopt cosmological parameters $q_0=0.5$, $H_0$ = 50 km s$^{-1}$ Mpc$^{-1}$, $\lambda_0=0$ throughout the paper for consistency with previous work unless noted.

\section{Observations}

  A deep high-resolution $K^\prime$-band image of the cluster RX J0848.9+4452 was obtained with the Subaru telescope equipped with {\it The Infrared Camera and Spectrograph } (IRCS) \citep{kob00} during the night of 5 January, 2001 (UT). The pixel scale was $0.058''$ per pixel, for a field of view of $1' \times 1'$. We accumulated 120-sec exposures with a 5-point dithering pattern for a total net integration time of 4 hours. The seeing was moderate and the FWHM of the stellar objects in the final combined image $\sim 0.''45$. Data were reduced in the standard manner; after dark subtraction and flat-fielding using the dome-flat frame, we subtracted a median sky frame from each image, derived from the five preceding and following frames. Magnitude calibration  to the Mauna Kea system $K$ band was carried out by observing the UKIRT Faint Standard Stars (Hawarden et al. 2001) observed before and after the cluster observation.  We ignored the $K^\prime-K$ color correction.
 
 In addition to the IRCS data, we also used a part of the Subaru Suprime-Cam image of the field \citep{nak02} and  archival WFPC2 data from Hubble Space Telescope (PID:8269, PI:Elston)  in order to study the galaxy properties at the optical wavelengths.

\section{The Brightest Cluster Galaxy in RX J0848.9+4452 }

  Figure 1 shows the $K'$-band image of the central region of the cluster. The side of the box is  $7''$, which corresponds to 58 kpc at $z=1.26$. In this image, the brightest galaxy previously catalogued as galaxy $\#1$ in \cite{ros99} is clearly resolved into two components.  The projected separation between the peaks of the two components is 0.$^{\prime\prime}$7 or 5.8 kpc assuming both galaxies lie at the cluster redshift. 

 Since the field is crowded, it is not straightforward to derive photometric properties for each galaxy. We modeled the galaxies in Figure 1 using a de Vaucoleurs' profile with centroids as determined from the SExtractor algorithm \citep{bea96}: the parameters using in our modeling were:  effective radius, effective surface brightness (or equivalently, total magnitude), axial ratio, and position angle for the five galaxies. Figure 2 shows artificial galaxy images built using the IRAF ARTDATA package using the best fitting model parameters (left-hand panel) and residuals after subtraction of the actual image (right-hand panel). We summarize the results of the fitting in Table 1. Galaxy B is the first ranked cluster galaxy whereas galaxy A is still about 0.5 magnitudes brighter than the next brightest spectroscopic member (\# 11) in \cite{ros99}

 Figure 2 shows both positive and negative residuals at about $2\%$ and $5\%$ of the total flux of galaxy B. Although our assumption that galaxies in Figure 1 are represented by de Vaucouleurs profiles is certainly an oversimplification, the errors introduced in the modeling are therefore too small to affect our discussion of the $K$-band Hubble diagram. In order to test this we smoothed our image to the $0.9''$ seeing and the resolution reported in \cite{ros99} and derived a luminosity within 50 kpc metric aperture for the combined galaxy of $K=16.84$ which compares well with the published value of $K=16.72$ (Rosati et al. 1999).

 Figure 3 shows the $K$-band Hubble diagram of the brightest cluster galaxies. The central galaxies  A and B in RX J0848.9+4452 are plotted as filled squares together with low and intermediate redshift data from \cite{com98} (as open circles) and tracks for passive and no-evolution models (cf. Figure 2 in Burke et al. 2000). Magnitudes ($K=17.9$ and 17.4 for galaxy A and B, respectively) are measured within 50 kpc metric apertures for consistency with previous data. Neither A nor B is as bright as the local brightest cluster galaxies evolved passively to $z=1.26$. This result holds for another cosmological model with $\Omega_m = 0.3$, $\Omega_\lambda = 0.7$, and $H_0 = 70$ km s$^{-1}$Mpc$^{-1}$ where the predicted $K$-band magnitude at $z=1.26$ of the models with same present-day luminosity is about 0.8 mag fainter than in Figure 3. Although RX J0848.9+4452 is an X-ray cluster and  a massive system in an evolved dynamical stage with high X-ray temperature, its `BCG' appears to be at an early evolutionary phase relative to its counterparts in the local universe.

\section{Discussion}

 We have found that the brightest galaxy in the rich X-ray cluster RX J0848.9+4452 at $z=1.26$ consists of two components and is therefore fainter than its local counterparts.   We consider here the colors of the two components in order to constrain their stellar populations. We used a 4800s WFPC2 $F814W$ exposure of this field taken by HST. Figure 4 shows a $4'' \times 4''$ region of this image including our galaxies. The two components are clearly identified.  We smoothed the $F814W$ image to same resolution as our IRCS data and measured $I-K$ colors of the two components through $0.6''$-diameter apertures centered on the $K^\prime$-band centroids. Both components have very red colors, $I_{F814W}-K$ = 3.27$\pm$0.07 and 3.62$\pm$0.08 (in $AB$ magnitude), respectively, but galaxy A is 0.3 mag bluer than B.

  Figure 5 shows the two-color diagram of $K^\prime$-selected galaxies in our IRCS field. Optical colors are from the Suprime-Cam study of this cluster by \cite{nak02}. Our IRCS image was degraded to match the resolution and seeing ($1''$) of the Suprime-Cam image. Colors were measured within 2$^{\prime\prime}$-diameter apertures. Galaxies A and B are not separated in the image, and we treated them as a single galaxy. We measured  $R-I$ and $I-K$ colors (in Cousins and UKIRT-MK system) of our galaxies. We also plotted passive evolution models for old cluster galaxies from \cite{kod97}. The composite brightest cluster galaxy (A+B) is very close to predictions of passive-evolution models for a $M_V =-23$ (at $z=0$) galaxy observed at z=1.26 whose predicted $K$-band magnitude is 16.4. It is also one of the reddest galaxies in the field. Thus galaxies A and B of the rich z=1.26 cluster have stellar populations quite consistent with prediction of the passive-evolution model for the most luminous old elliptical galaxies, but are less massive than today's BCG's, having sizes comparable to those of normal ellipticals.

 If we assume that the two galaxies lie at the same redshift, their separation is as small as 6 kpc. Numerical simulations \citep{hau78,rol79} show that these objects should merge on timescales of less than a few $10^8$ years. We therefore argue that we are witnessing the major merger of two massive ($ \sim L^*$) elliptical galaxies to fuel the hierarchical growth of the BCG in RX J0848.9+4453 although the exact redshifts of the two systems and their relative velocity are necessary to draw definite conclusions on this issue. This result is consistent with models of hierarchical merging growth of BCGs. \cite{aes98} argued that the typical mass of BCGs at $z=1$ is expected to be about half of the present. \cite{dub98} also showed in his numerical simulation that the central galaxy forms through the merger of several massive galaxies in early history of cluster formation.

\cite{yam00} have measured $K$ luminosities of BCG's in three other $z > 1$ clusters: 3C324, B2 1335+28 and ClG J0848+4453 and found these galaxies are all fainter than the predictions of \cite{bur00} based on passive evolution models. These three clusters, however, are less X-ray luminous than RX J0848.9+4453 \citep{dic97,sta01} and/or are more irregular and less concentrated in their galaxy distributions \citep{kaj00,tan00,nak01,sta01}: the present result shows that BCG's are consistently less massive than present-day systems even in well-virialized systems.

It is not uncommon for present-day BCG's to contain multiple nuclei: about half of the BCG's in Abell clusters have multiple nuclei within 19.2 kpc \citep{hoe85,ton85}; however, objects with small separation are somewhat rare: only 7\% of the galaxies have multiple nuclei within 5 kpc of each other; the magnitude difference between the nuclei is larger than 1 magnitude in 60\% of the low-redshift sample of \cite{hoe85}. Therefore, the nascent BCG in RX J0848.9+4453 is a relatively rare object and may be an example of an equal mass merger as is predicted to occur at the top of the merging hierarchy.

 The two component galaxies are aligned to point roughly in the direction of the nearby $z=1.27$ cluster ClG J0848+4453 \citep{sta97} and are aligned with the outer X-ray contour of RX J0848.9+4452 \citep{sta01} This may be the high redshift equivalent of the alignment effect of BCGs in nearby clusters \citep{bin82,ful99}. Galaxies in the $z=1.2$ cluster around the radio galaxy 3C324 may be similarly aligned \cite{nak01}. \cite{wes94} proposed that this arises as clusters form by collimated infall along filaments. \cite{dub98} also showed in his N-body simulation that the shape and orientation of the BCGs become nearly congruent with the galaxy distribution in the host cluster. The observed evidence of the alignment effect at high redshift, if confirmed, implies that such collimated structure formation indeed took place at $z > 1$. 

\vspace{0.5cm}

  This work is based on data collected at Subaru Telescope, which is operated by the National Astronomical Observatory of Japan. This work is also based in part on observations with the NASA/ESA Hubble Space Telescope, obtained from the data archive at the Space Telescope Science Institute, U.S.A., which is operated by AURA, Inc.\ under NASA contract NAS5--26555. The Image Reduction and Analysis Facility (IRAF) used in this paper is distributed by National Optical Astronomy Observatories. U.S.A., operated by the Association of Universities for Research in Astronomy, Inc., under contact to the U.S.A. National Science Foundation.TK acknowledges JSPS postdoctoral research fellowship and the Daiwa-Adrian Prize 2001 for financial support.

\clearpage


\begin{figure}
\plotone{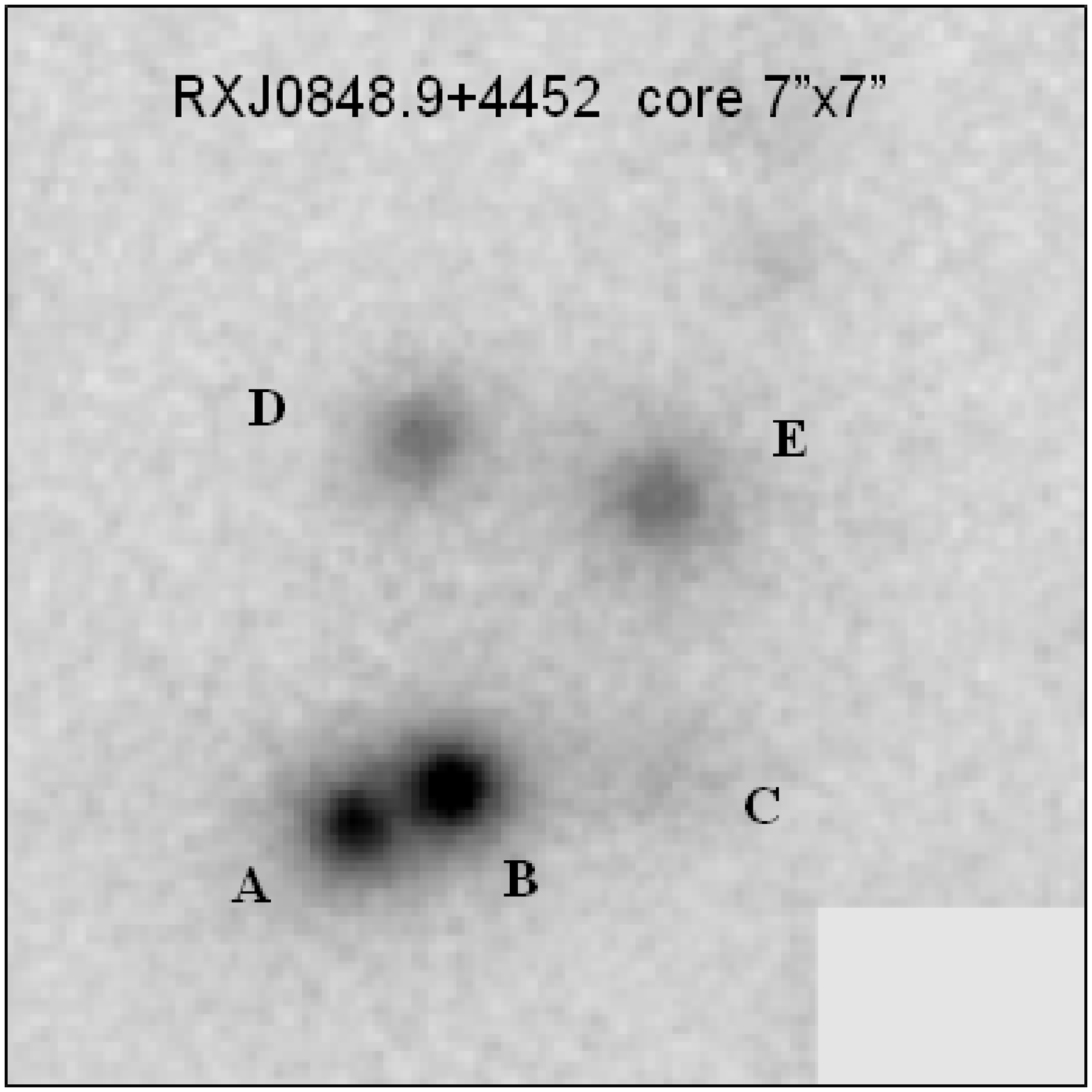}
\caption{The IRCS $K^\prime$-band image of the central part of the cluster RX J0848.9+4452 is shown. The side is 7$^{\prime\prime}$, which corresponds to 58 kpc at z=1.26. The north is up.}\label{fig:IRCS}
\end{figure}

\begin{figure}
\plotone{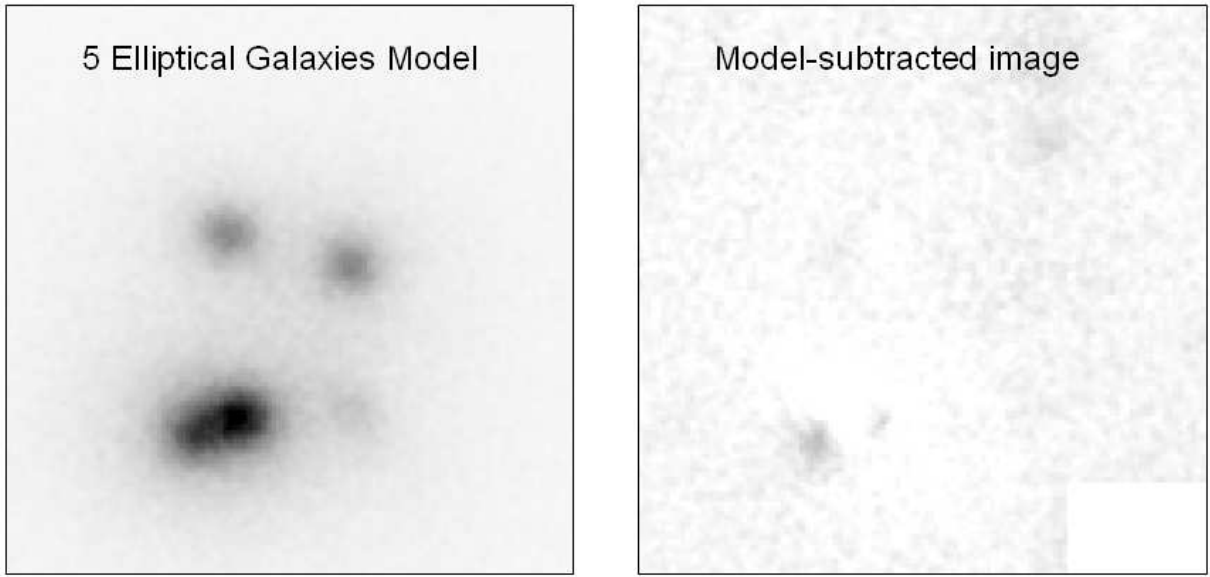}
\caption{The model image of five de Vaucouleurs-profile galaxies is shown in the left panel. The residual image after subtracting the model image from the observed one is shown in the right panel.}\label{fig:model}
\end{figure}
 
\begin{figure}
\plotone{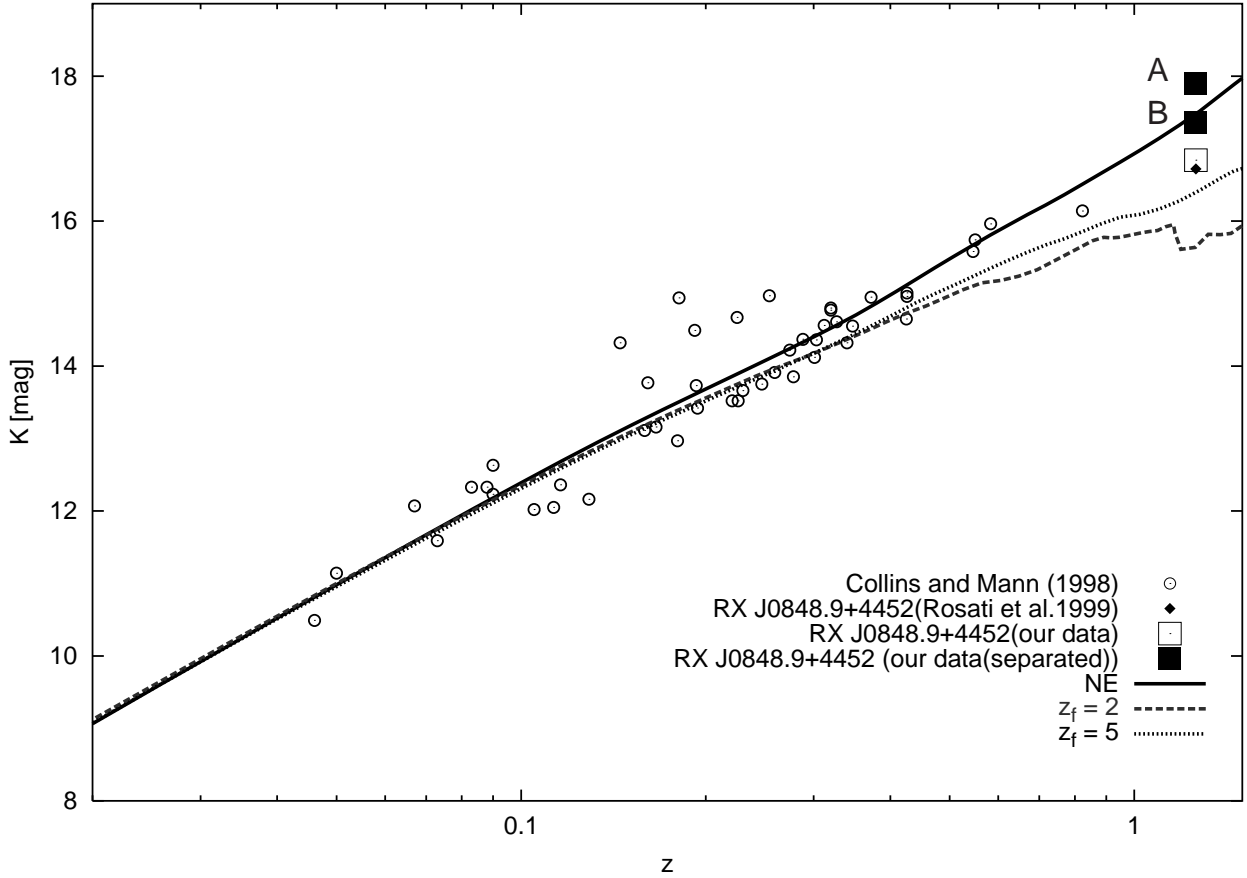}
\caption{The $K$-band Hubble diagram for the brightest cluster galaxies. The 50-kpc-diameter aperture magnitude for the galaxy A and B measured on the each model image are shown by the filled squares. The magnitude of the composite model galaxy (A+B) measured on the image smoothed to be matched with the seeing of Rosati et al. (1999) is also plotted by the open square. The data for the low- and intermediate-redshift cluster galaxies taken from Collins and Mann (1998) are plotted by the open circles. The magnitude value, $K=16.72$, quoted from Rosati et al. (1999) is also plotted (filled diamond). Lines show the model predictions for no evolution (solid) and passive-evolution with $z_f = 2$ (dashed) and $z_f = 5$ (dotted).}\label{fig:Hubble}
\end{figure}

\begin{figure}
\plotone{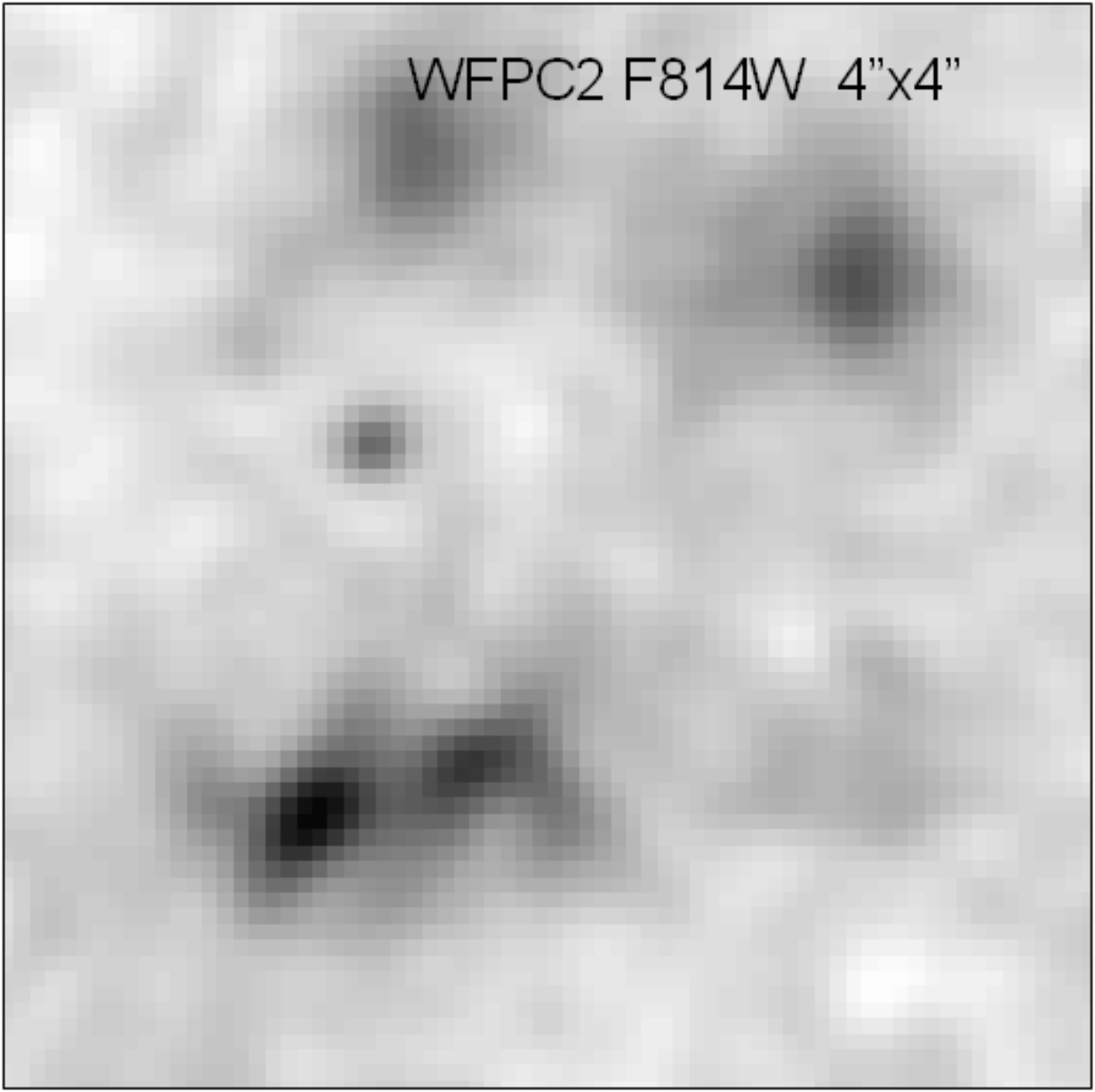}
\caption{The HST WFPC2 $F814W$-band image of the very central part of the cluster RX J0848.9+4452 is shown. The side is 4$^{\prime\prime}$ and the north is up.}\label{fig:WFPC}
\end{figure}

\begin{figure}
\plotone{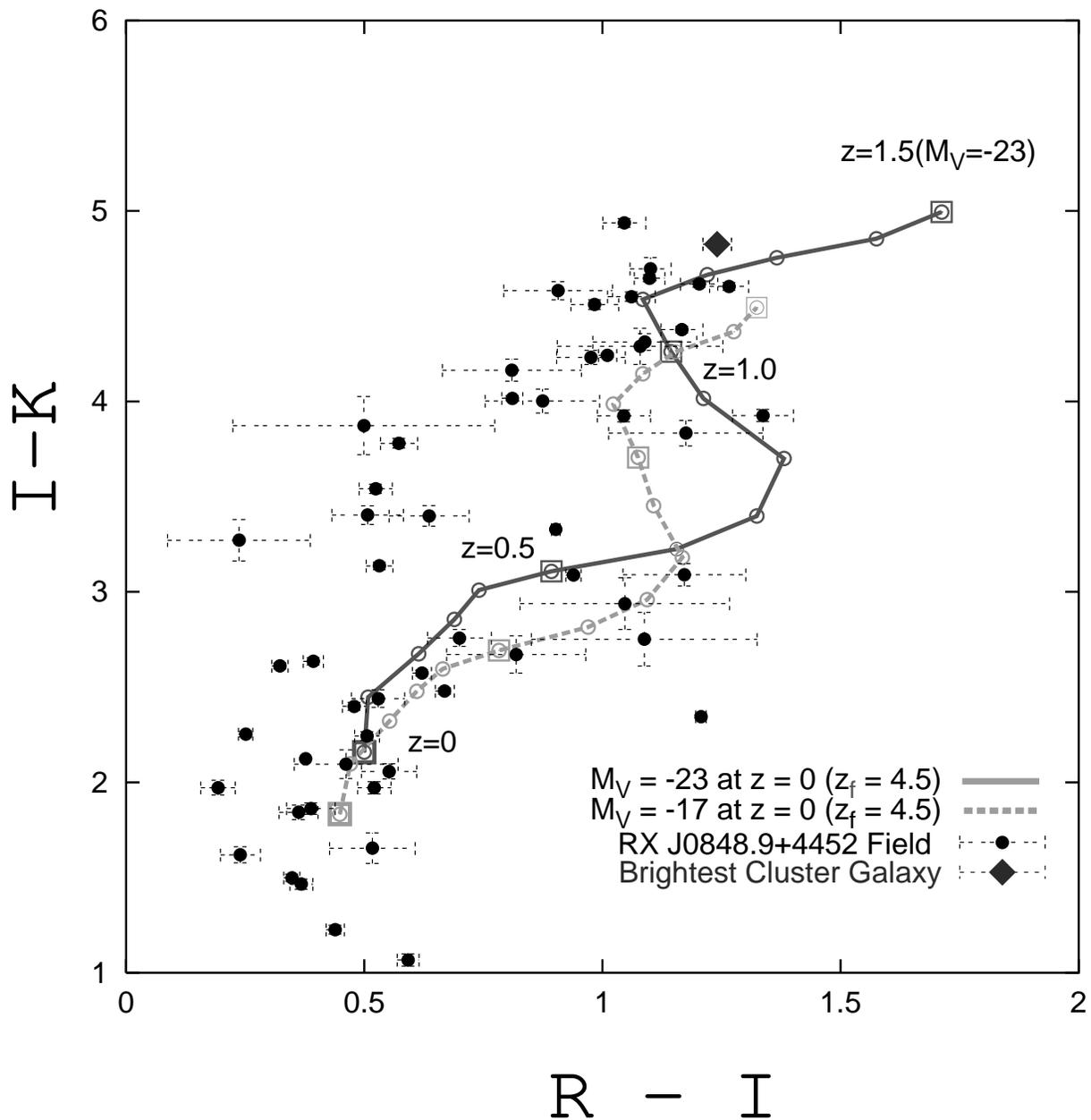}
\caption{The two color diagram for the $K^\prime$-selected galaxies in the 1$^\prime$$\times$1$^\prime$ field of the cluster RX J0848.9+4452. The data point for the unresolved BCG in RX J0848.9+4452 is shown by the filled diamond. For the guide, we also plotted the passive evolution models of old cluster galaxies with $M_V=-23$ and $M_V=-18$ (at z=0) observed at various redshift from Kodama \& Arimoto (1997). }\label{fig:twocolor}
\end{figure}

\clearpage

\begin{deluxetable}{ccccc}
\tabletypesize{\scriptsize}
\tablecaption{The best-fitted parameters of the model galaxies. \label{tbl-1}}
\tablewidth{0pt}
\tablehead{
\colhead{galaxy} & \colhead{$K_{\rm total}$}   & \colhead{ r$_{\rm e}$}   &
\colhead{ellipticity} & \colhead{P.A.}\\
\cline{1-5}
\colhead{      } & \colhead{mag.}   & \colhead{arcsec.}   &
\colhead{b/a} & \colhead{deg.} 
}
\startdata
 A &  17.5 &  1.4 &   0.95 &   94 \\
 B &  17.0 &  1.7 &   0.74 &   86 \\
 C &  19.6 &  1.2 &   0.48 &   37 \\
 D &  18.0 &  1.3 &   0.79 &   74 \\
 E &  17.8 &  1.6 &   0.97 &  141 \\
\enddata
%
%
\end{deluxetable}

\end{document}